\documentclass[twocolumn,superscriptaddress,english,prl,showpacs,longbibliography]{revtex4-1}
\usepackage[colorlinks=true,urlcolor=blue,citecolor=blue,linkcolor=blue]{hyperref}

\usepackage{amsmath}
\usepackage{graphicx}% Include figure files
\usepackage{bm}% bold math
\usepackage{hyperref}% add hypertext capabilities
\usepackage{color}
\usepackage{amssymb}
\usepackage{graphicx}
\usepackage{color}
\usepackage{mathrsfs}
\usepackage{float}
\usepackage{indentfirst}
\usepackage{txfonts}
\usepackage{algorithm}  
\usepackage{algpseudocode}  
\usepackage{balance}
\usepackage{flushend}
\newcommand{\s}{\mathbf {s}}

\newcommand{\argmin}{\mathop{\mathrm{argmin}}\limits}
\begin{document}

\preprint{APS/123-QED}

\title{Contracting Arbitrary Tensor Networks: General Approximate Algorithm and \\ Applications in Graphical Models and Quantum Circuit Simulations}% Force line breaks with \\

\author{Feng Pan}
\thanks{These authors contributed equally.}
\affiliation{
 CAS Key Laboratory for Theoretical Physics, Institute of Theoretical Physics, Chinese Academy of Sciences, Beijing 100190, China
}
\affiliation{
 School of Physical Sciences, University of Chinese Academy of Sciences, Beijing 100049, China
}
\author{Pengfei Zhou}
\thanks{These authors contributed equally.}
\affiliation{
 CAS Key Laboratory for Theoretical Physics, Institute of Theoretical Physics, Chinese Academy of Sciences, Beijing 100190, China
}
\affiliation{
 School of Physical Sciences, University of Chinese Academy of Sciences, Beijing 100049, China
}

\author{Sujie Li}
\thanks{These authors contributed equally.}
\affiliation{
 CAS Key Laboratory for Theoretical Physics, Institute of Theoretical Physics, Chinese Academy of Sciences, Beijing 100190, China
}
\affiliation{
 School of Physical Sciences, University of Chinese Academy of Sciences, Beijing 100049, China
}
\author{Pan Zhang}
\email{panzhang@itp.ac.cn}
\affiliation{
 CAS Key Laboratory for Theoretical Physics, Institute of Theoretical Physics, Chinese Academy of Sciences, Beijing 100190, China
}
\affiliation{
	 School of Fundamental Physics and Mathematical Sciences, Hangzhou Institute for Advanced Study, UCAS, Hangzhou 310024, China
}
\affiliation{
	International Centre for Theoretical Physics Asia-Pacific, Beijing/Hangzhou, China
}

\date{June 30 2020}% It is always \today, today,

\begin{abstract}
%Tensor networks are powerful tools for solving quantum many-body problems, which are usually defined on lattices where efficient contraction algorithms exist. However, when applied to problems out of physics such as optimization, inference, and learning and optimization problems, the underlying network connections are usually far from lattices.
%, and there is no existing general algorithm that can approximately contract every network efficiently. 
We present a general method for approximately contracting tensor networks with an arbitrary connectivity. This enables us to release the computational power of tensor networks to wide use in inference and learning problems defined on general graphs. We show applications of our algorithm in graphical models, specifically on estimating free energy of spin glasses defined on various of graphs, where our method largely outperforms existing algorithms including the mean-field methods and the recently proposed neural-network-based methods. We further apply our method to the simulation of random quantum circuits, and demonstrate that, with a trade off of negligible truncation errors, our method is able to simulate large quantum circuits that are out of reach of the state-of-the-art simulation methods.
%using super computers. 
%while produces only negligible truncation errors.
%it displays significant speed up and memory efficiency over the state-of-the art exact simulations while produces only negligible truncation errors. 
% \begin{description}
% Secondary publications and  information retrieval purposes.
% You may use the \texttt{description} environment to structure your abstract;
% use the optional argument of the \verb+\item+ command to give the category of each item. 
% \end{description}
\end{abstract}

\maketitle

As a powerful method to alleviate the "curse of dimensionality" in high-dimensional modeling and data analysis, the tensor networks find wide applications in many areas of science and technology.
%including quantum many body physics, statistical mechanics, data science, and machine learning, etc.
In quantum many-body physics, tensor networks on lattices including the matrix product states (MPS)~\cite{vidal2004efficient,verstraete2004matrix}, and the projected entangled pair states (PEPS)~\cite{verstraete2004renormalization} have great success in the study of strongly correlated systems; in statistical mechanics, calculation of the partition function can be naturally converted to a tensor network contraction problem~\cite{levin2007tensor}; in computer science, the number of solutions of constraint satisfaction problems can be computed via tensor networks~\cite{kourtis2018fast}; in data science, tensor networks and tensor decompositions are important tools for data compression and dimensionality reduction~\cite{cichocki2016tensor}. Recently, tensor network methods have been successfully extended to machine learning, in compressing a neural network~\cite{gao2019compressing}, giving an efficient image classifier~\cite{stoudenmire2016supervised}, and working as generative models in the unsupervised learning~\cite{han2018unsupervised,cheng2019tree}.

%Tensor networks are important tools in the study of quantum many-body physics. For one-dimensional systems, the matrix product states (MPS) ~\cite{vidal2004efficient,verstraete2004matrix} and the density matrix renormalization group (DMRG)~\cite{white1992density} have almost dominated the numerical studies; in two-dimensional systems, 2-D tensor networks such as PEPS~\cite{verstraete2004renormalization} and PESS give the state-of-the-art results; in critical systems, tensor networks such as MERA ~\cite{PhysRevLett.99.220405,PhysRevLett.101.110501,PhysRevLett.115.200401} also play an important rule. 
%Analogous to quantum many body physics, in many problems %of science and technology, the 
%In addition to quantum physics, tensor networks also have wide applications in many fields of science and engineering. For example the statistical mechanics problems such as computing the partition function and marginals can be naturally converted to a tensor network contraction problem~\cite{levin2007tensor}; in data  science, tensor networks become a powerful tool for dimensionality reduction~\cite{cichocki2016tensor}; recently, tensor network methods have been successfully extended to machine learning, in compressing a neural network~\cite{gao2019compressing}, giving an efficient image classifier~\cite{stoudenmire2016supervised}, and working as generative models for unsupervised learning~\cite{han2018unsupervised,cheng2019tree}.

\begin{figure}[tb]
\centering
\includegraphics[width=\columnwidth]{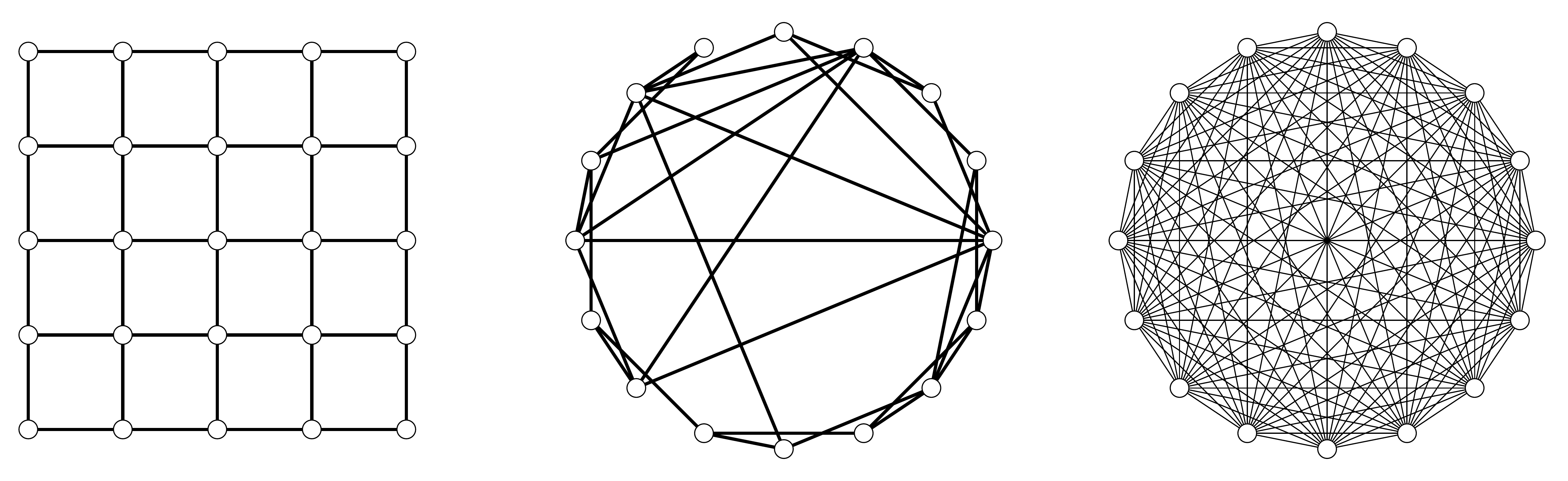}
\includegraphics[width=.99\columnwidth]{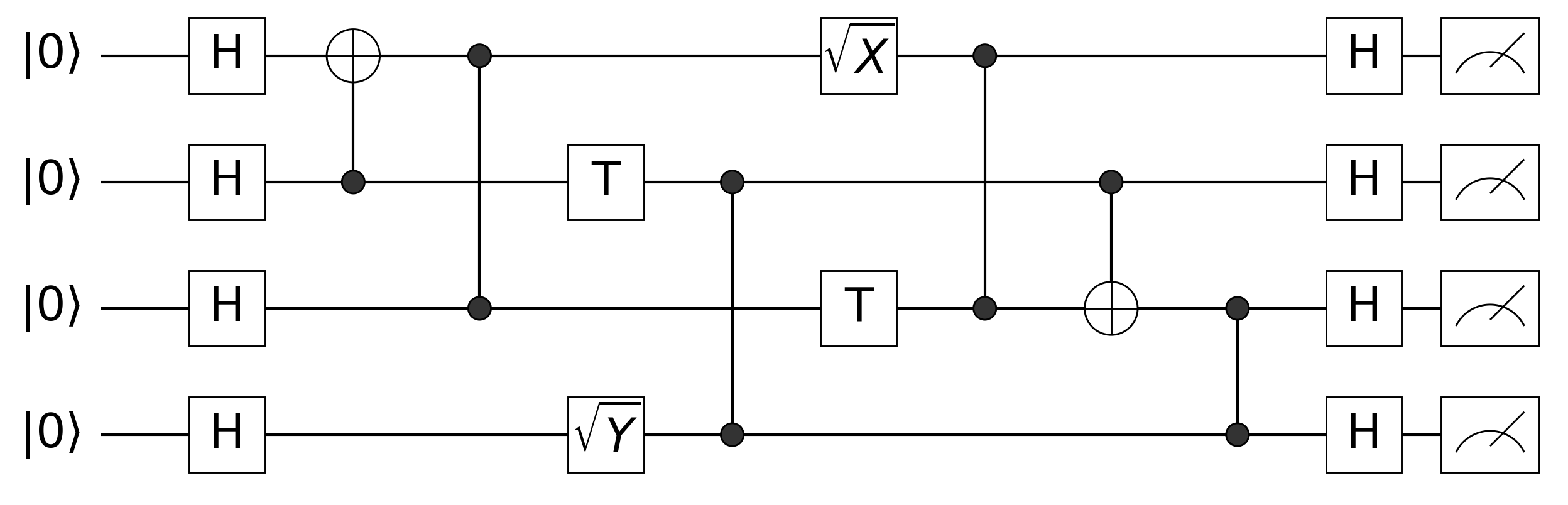}
\caption{\label{fig:tn-graph} 
Illustration of connectivity graph of the tensor networks we aim to contract: two-dimensional lattices, random graphs, fully connected graphs, and those defined by the quantum circuits.
%From left to right, tensor networks representations of partition function of two-dimensional Ising model, Watts-Strogatz model~\cite{watts1998collective} with average degree $4$ and rewiring probability $0.4$ and full connected Sherrington-Kirkpatrick model~\cite{sherrington1975solvable}
\label{fig:tns}}
\end{figure}

Despite its wide use, however, the capability of the tensor networks is so far limited to either small-dimensional systems where the exact contraction is tractable, or high-dimensional systems only on regular lattices with local interactions, where there exist efficient contraction algorithms, e.g., the renormalization group~\cite{white1992density,levin2007tensor,PhysRevB.86.045139,adachi2019anisotropic} and the block decimation~\cite{PhysRevB.78.155117}. On general systems with long range interactions and irregular connectivity (such as the graphs depicted in Fig.~\ref{fig:tns}), the tensor network method is rarely applied, due to intractability of efficient contraction: to the best of our knowledge, there is no general method that exists for approximately contracting arbitrary tensor networks. This sets limitations on applying tensor networks to many areas, such as graphical models, statistical inference, and machine learning problems.

In this Letter we aim to break this limitation. We propose a general method for approximately contracting tensor networks on an arbitrary graph, based on a method we term as "MPS calculus": the initial and intermediate tensors produced during the tensor contractions are represented, compressed, and operated using the matrix product states in the canonical form. This allows us to deal with large intermediate tensors, which cannot be stored in the memory in its original form. During the contraction process we iteratively detect low-rank structures and apply low-rank approximations to reduce computational complexities of the contraction, using approaches analogous to the density matrix renormalization group (DMRG)~\cite{white1992density}, until the final result, a scalar $Z$, is obtained. We show applications of our method in graphical models, where $Z$ represents the normalization factor of the joint distribution of a large number of random variables (i.e., the partition function in physics), and applications in quantum circuit simulations where $Z$ represents a single amplitude of the quantum circuit.
%To the best of our knowledge, contracting arbitrary tensor networks taking the matrix product states as building blocks has not been explored before. 
%In fact we are not aware of any algorithm which is able to contract arbitrary tensor networks approximately. 

\paragraph{\label{sec:contract}Contracting arbitrary tensor network.---}
Our method relies on two ideas: (1) representing every tensor in the network by a matrix product state in the canonical form and (2) performing low-rank approximations based on the MPS representations during contraction. The matrix product state, also known as the tensor train in mathematics~\cite{oseledets2011tensor}, is a one-dimensional tensor network composed of three-way tensors (and matrices in the boundary). 
%The MPS has been widely applied to quantum many body problems, especially in one-dimension. 
A straightforward advantage of MPS is the parameter efficiency: an $n$-way tensor $\mathcal{ A}\in\mathbb {C}^{d^n}$ can be represented by an MPS of virtual bond dimension $\chi$ with only $(n-2)d\chi^2+2d\chi$ parameters, using e.g., the DMRG~\cite{white1992density}. With a large enough $\chi$, the MPS can faithfully represent the original tensor and hence give an exact result. With limited computational resources, one would restrict the bond dimensions, performed as an approximation to the underlying raw tensor $\mathcal {A}$.
%and introduce a certain amount of error, depending on how many intrinsic low-rank structures there are in the underlying raw tensor $\mathcal {A}$.
%A simple example of the converting is depicted in Fig.~\ref{fig:contraction} where the original $5$-way tensors are converted to MPSes. 
%The amount of errors introduced with a finite bond dimension depend  there are low-rank structures in the underlying raw tensor $\mathcal {A}$. 
%In the quantum many-body systems with a local Hamiltonian, the ground-state usually has relatively low entanglements, satisfying the \textit{area law}, hence the MPS representation usually gives a good approximation.
Another characteristic of MPS is the canonical form, which can be achieved using QR decompositions or singular value decompositions~\cite{schollwock2011density,orus2014practical}. The first advantage of the canonical form is fixing the gauge degree of freedom, which eliminates the nonuniqueness in representing a raw tensor. More importantly, in the canonical form, the sum of discarded squared singular values corresponds to the loss of $\mathcal L_2$ norm of the whole MPS, rather than the local three-way tensor, which allows low-rank approximations on a global scope. 
%As an advantage of reducing the length of formulas, we will use tensor diagram notations whenever possible.

Given a tensor network composed of tensors $\mathcal{A}^{(1)}$... $\mathcal{A}^{(n)}$ and edges connecting the tensors, the high-level desciption of our algorithm, MPS calculus, is processed as follows: (1) Convert every tensor to a MPS. (2) If there are no edges left, return; else select an edge $(ij)$ according to a contraction order. (3) "Contract" ${\mathbf{\mathcal{A}^{(i)}}}$ and $\mathbf{ \mathcal {A}^{(j)}}$, store as ${\mathbf{\mathcal{A}^{(i)}}}$; delete ${\mathbf{\mathcal{A}^{(j)}}}$. (4) If ${\mathbf{\mathcal{A}^{(i)}}}$ connects to ${\mathbf{\mathcal{A}^{(k)}}}$ by two edges, "merge" the edges to a single edge using "swap" operations and low-rank approximations with singular value decomposition (SVD); then go to step 2.

\begin{figure}[tb]
	\centering
	\includegraphics[width=\columnwidth]{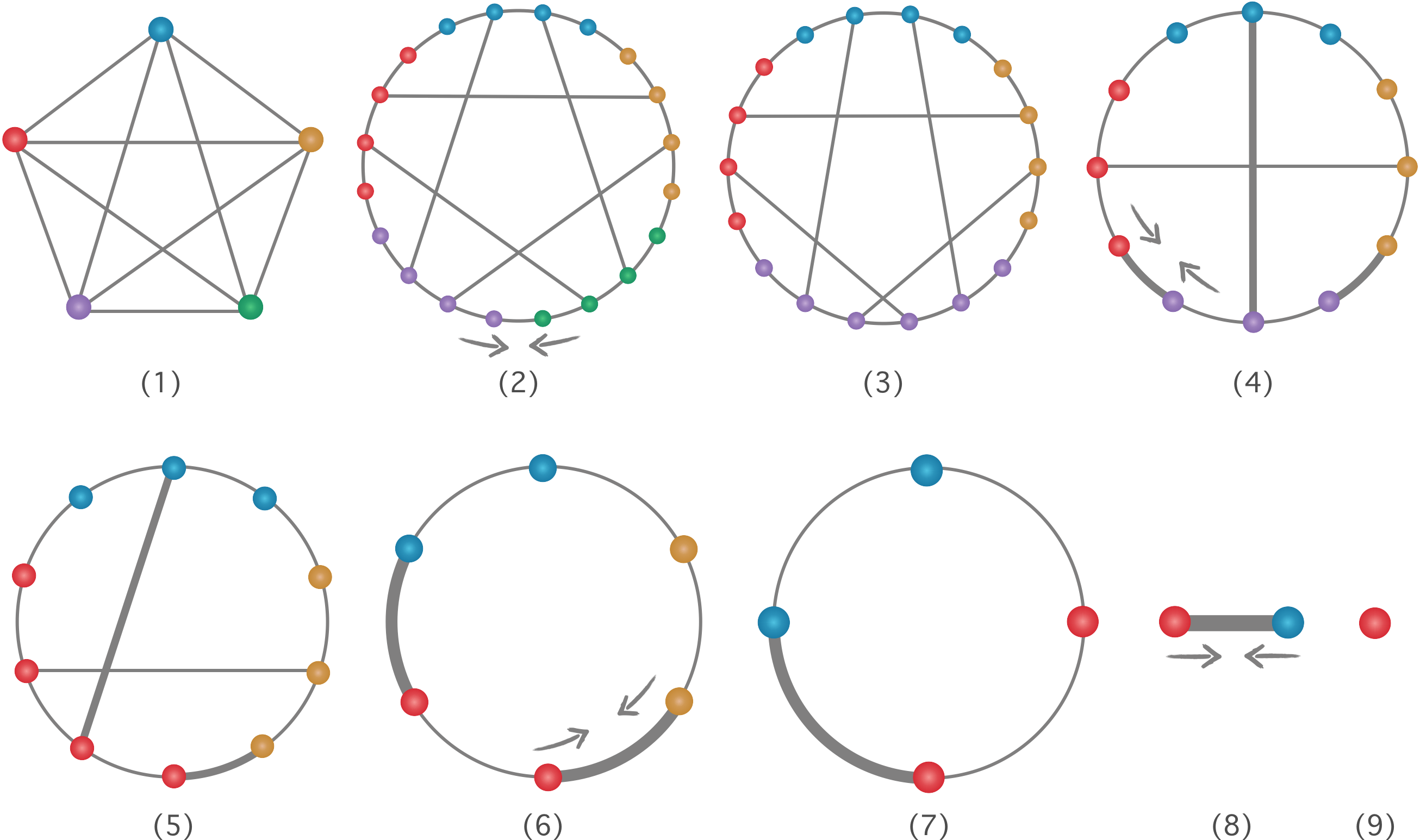}
	\caption{\label{fig:contraction} Pictorial representation of our algorithm in contracting a tensor network with five tensors; see descriptions in main text. }
	%In step (2), every tensor appears in (1) is converted to a MPS in the canonical form. During step (3) to step (8), edges of the tensor network are contracted one by one, using the \textit{MPS calculus}, finally producing a scalar in (9).}
\end{figure}

A pictorial representation of the algorithm is sketched in Fig.~\ref{fig:contraction} using a simple example of contracting a fully connected tensor network with five tensors, as shown in panel (1). In panel 2, every tensor that appears in 1 is converted to a MPS in the canonical form; During step 3-8, edges of the tensor network are contracted one by one, finally producing a scalar in step 9. For further details about the algorithm and order choices, please refer to the Supplemental Material and Refs~\cite{orus2014practical,schollwock2011density,white1992density,levin2007tensor,xie2009second,PhysRevB.86.045139,markov2008simulating,dumitrescu2018benchmarking,chen2018classical,boixo2017simulation,PhysRevLett.123.190501,gray2020hyper,huang2020classical}.

\begin{figure}[tb]
	\centering
	\includegraphics[width=\columnwidth]{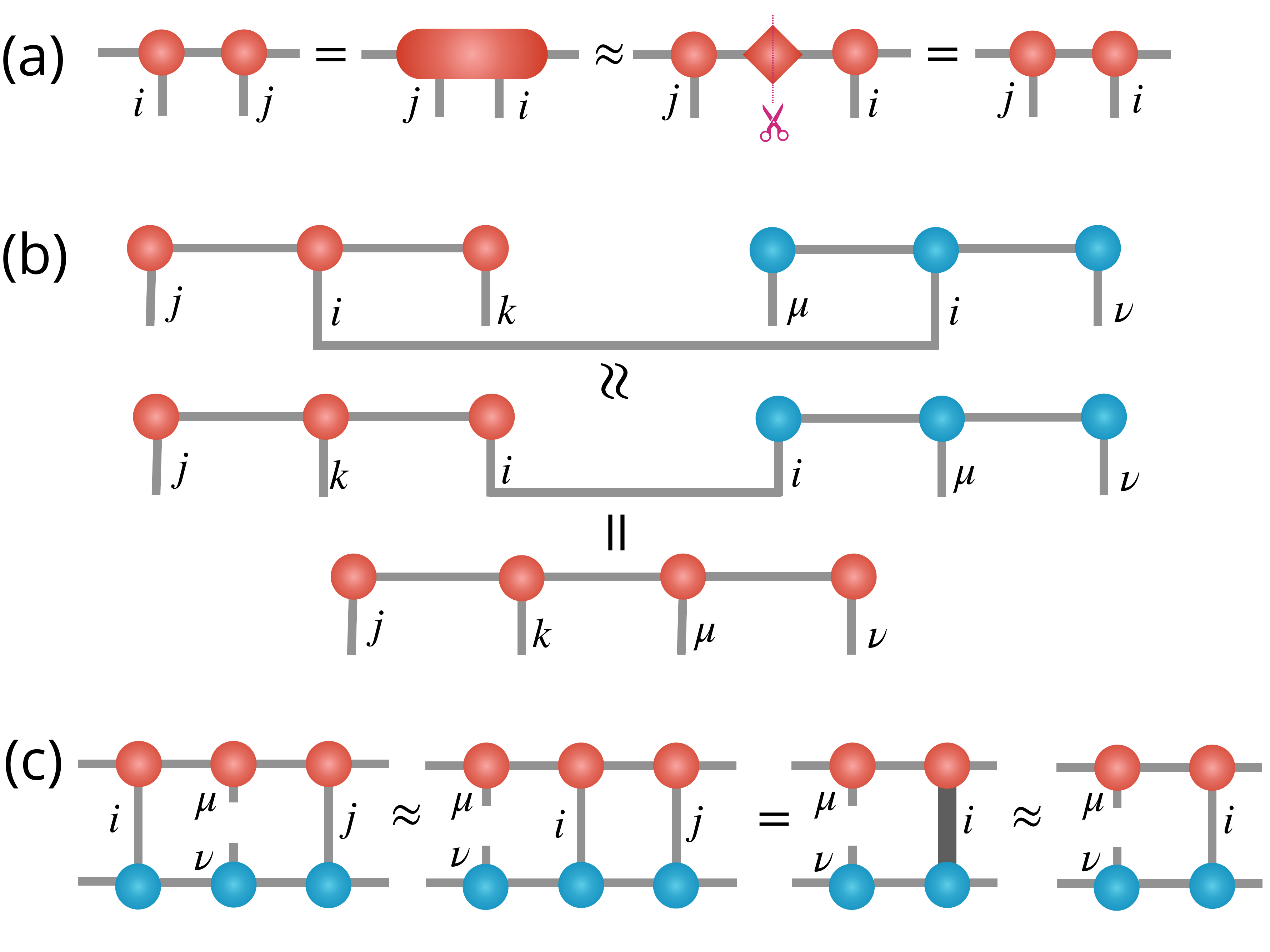}
	\caption{\label{fig:eat} Illustration of the (a) swap, (b) contract, and (c) merge operations. The scissor symbol indicates truncation of the singular values.}
\end{figure}

The contract operation is processed by merging two tensors to a single tensor by summing over the common index (say $i$) of them. Since all of them are MPSes, we need to move the common index $i$ to the tail of the first tensor and to the head of the second tensor, using the swap operations. The swap operation switches the positions of two indices in the original tensor, by swapping two adjacent tensors in the mps, with a similar functionality as the swap gate in the quantum information. This operation increases entanglements of the MPS, and the maximum bond dimension could increase to $d\chi$, where $\chi$ denotes the virtual bond dimension of the MPS and $d$ is the dimension of the physical indices. If $d\chi$ is greater than $\widehat \chi$, the preset limit on the virtual bond dimension, we canonicalize the MPS then truncate the bond dimension to $\widehat \chi$ during the singular value decomposition. An example of swap and contract are illustrated using tensor diagram notations in Fig.~\ref{fig:eat}, where the scissor symbol indicates truncating of the dimension in the diagonal matrix.

After the contraction, the obtained tensor could have two indices, say $j$ (with bond dimension $d_j$) and $k$ (with bond dimension $d_k$) linked together to another tensor, due to existence of a triangle with three end tensors. In this case, we move indices $j$ and $k$ to adjacent positions using the swap operations and merge the two corresponding tensors to a three-way tensor with a larger physical bond dimension $d_jd_k$. If it exceeds $\widehat D$, the preset maximum physical bond dimension, we canonicalize both tensors, then do SVD together with a truncation on singular values to reduce the bond dimension from $d_jd_k$ to $\widehat D$. The process is illustrated in Fig.~\ref{fig:eat} (c).

The operations swap, contraction, and merge are repeated until the overall tensor network is finally contracted to a scalar $Z$. Our algorithm takes two parameters, the  maximum physical bond dimension $\widehat D$ and the maximum virtual bond dimension $\widehat\chi$ of the MPSes. The space complexity of the algorithm is bounded above by $\mathcal {O}(\widehat{D}\widehat\chi^2)$, and the time complexity is dominated by singular value decompositions adopted in the swap operations, which is $
\mathcal {O}(\widehat{D}^3\widehat\chi^3)$. Apparently, it is a polynomial algorithm that is able to contract arbitrary tensor networks with a limited amount of computational resources. Moreover, our method enjoys an efficient approximation scheme analogous to the DMRG method, which allows dynamically adjusting dimensions of the tensors. In the following text we will give applications of our algorithms, the inference and learning in the graphical models, and the simulation of quantum circuits, to empirically evaluate our method.

We noticed that in~\cite{jermyn2020automatic} the authors have proposed a general tensor network contraction algorithm by representing large intermediate tensors using the tree tensor network and reducing loop length using local singular value decompositions. Compared with~\cite{jermyn2020automatic}, our method is capable of using larger bound dimension because the MPS has lower space complexity than the tree tensor network. Moreover, the canonical form of MPS allows more effective approximations.

\paragraph{\label{sec:gm} Applications to graphical models--}

Graphical models are important tools for representing joint probability distributions over a large number of random variables that interact with each other, and find important applications in many fields in science and engineering. Without loss of generality, in this Letter we use the classic example of the graphical model, the Ising model and spin glasses in the statistical physics to demonstrate the power of our method. In this problem, the joint probability of $n$ spins $\s \in \{\pm 1\}^n$ follows the Boltzmann distribution
%\begin{equation} P(\s) = \frac{1}{Z}\exp({-\beta E(\s)}), \label{eq:boltzmann} \end{equation}
$P(\s) = \frac{1}{Z}\exp{[-\beta E(s)]}$,
where $E(\s)$ is the energy function of a configuration $\s$, $\beta$ is the inverse temperature and $Z$ is the partition function. 
Given a problem instance, an essential problem is computing the free energy $F = -\frac{1}{\beta} \ln Z$. 
%Based on an accurate estimation of the free energy, the observables such as magnetizations and correlations, even the unbiased samples, can be obtained based on the free energy estimate.
%, compute macroscopic properties of the system such as magnetizations and correlations, and how to sample from the Boltzmann distribution efficiently. i
However, this problem belongs to the class of $\#P$ problems hence it is hopeless to find polynomial algorithms for solving it exactly.
In physics, many approximate algorithms have been developed. These include Markov chain Monte Carlo methods~\cite{PhysRevLett.86.2050} and mean-field methods that parametrize a variational distribution by minimizing the variational free energy. Recently in ~\cite{wu2019solving}, the mean-field methods have been extended by employing the autoregressive neural networks as a variational distribution, which, in principle has a strong expressive power.

Any probability distribution over discrete variables is a tensor, thus every graphical model can be converted to a tensor network by introducing copy tensors on each node of the graph, and matrices (or tensors) on each edge (or multibody factor) of the (factor) graph. The computation of the partition function $Z$ naturally translates to contraction of the tensor network defined exactly on the same graph. As an example, consider the celebrated pairwise Ising spin glass model with $n$ variables: its energy function is defined as $E(\s)=-\sum_{(ij)\in\mathcal E}J_{ij}s_is_j$, with $\mathcal E$ denoting a set of edges and $J_{ij}$ denoting couplings between two spins $i$ and $j$. The partition function can be written formally as

\begin{equation}
\label{eq:z} 
Z=\sum_\mathbf{s}\prod_{(ij)\in\mathcal E}e^{\beta J_{ij}s_is_j}=\mathbf{Tr}\left( \mathcal{A}^{(1)} \times \mathcal{A}^{(2)}\times\cdots\times \mathcal{A}^{(n)}\right), \end{equation}
where the symbol $\times$ represents contraction of tensors $\{\mathcal{ A}^{(i)}\}$, each of which is given by contracting a copy tensor with matrices defined on the edges connected to node $i$,
$$
\mathcal{A}^{(i)} = \mathbf{ \mathcal{I}}_{d_i\times d_i}\times{\mathbf{B}_{j\in\partial i}}\times{\mathbf{B}_{k\in\partial i}}\times\cdots\times{\mathbf{B}_{l\in\partial i}}.
$$
Here $\mathbf{\mathcal{I}}_{d_i\times d_i}$ is a copy tensor, i.e. a diagonal tensor with order equal to the degree (number of neighbors) $d_i$ of node $i$, with one on the diagonal entries and zero on the other entries. $\partial i$ denotes the set of neighbors of node $i$, and the matrix
$\mathbf B_{j\in\partial i}$ is a $2\times 2$ matrix with $[\cosh(\beta J_{ij})/2]^{1/2}+[\sinh(\beta J_{ij})/2]^{1/2}$ on the diagonal and $[\cosh(\beta J_{ij})/2]^{1/2}-[\sinh(\beta J_{ij})/2]^{1/2}$ on the off-diagonal entries.

 \begin{figure*}[tb]
	\includegraphics[width=2\columnwidth]{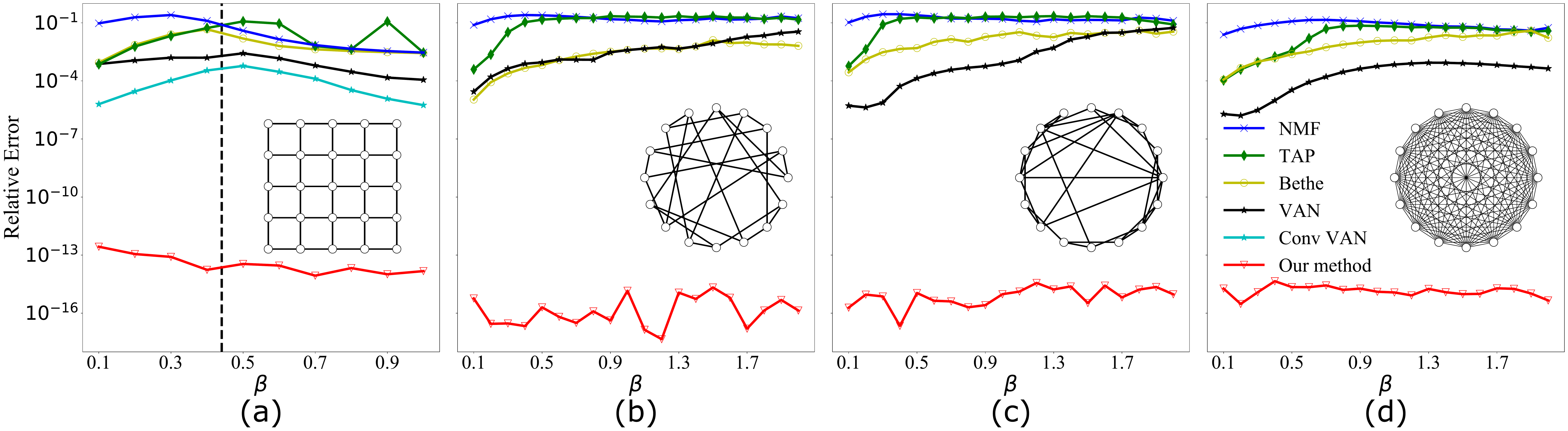}
	\caption{\label{fig:relative error} Relative errors of the free energy to exact solutions obtained by different methods on various models. Insets: illustrations of the underlying connectivity graph with smaller sizes. (a) Ferromagnetic Ising model on a $16\times 16$ square lattice; the exact solutions are given by~\cite{kac1952combinatorial}, and the vertical dashed line represents the phase transition of an infinite system.
		(b) Ising spin glass model on random regular graphs of $80$ nodes with degree $k=3$; couplings $J_{ij}$ are drawn from normal distribution with zero mean and unit variance. (c) Ising spin glass model on the Watts-Strogatz graphs of $70$ nodes with average degree $c=4$ and rewiring probability $p=0.4$. The exact solutions are given by enumerating all configurations of feedback set of graphs~\cite{pan2019solving}. (d) The Sherrington-Kirkpatrick model with $n=20$ spins; exact solutions are given by enumerating $2^n$ configurations. Data points are averaged over $10$ random instances.}
\end{figure*}

After converting the graphical model to tensor network, our method directly applies to computing free energy of the problem defined on arbitrary graphs. Observe that our algorithm is exact when the graph is a tree, because, by minimizing the size of the intermediate tensor, it performs variable eliminations iteratively on leaves of the tree and hence reduces to the belief propagation algorithm. On other graphs, our algorithm might generate truncation error $\epsilon_{\textrm{SVD}}$. Empirically we observe that the error $\epsilon_{\textrm{SVD}}$ is several magnitudes smaller than the error of the obtained free energy $\epsilon_{\textrm{F}}$, but so far it is not clear to us how to relate the two errors analytically. We subject to numerical experiments to demonstrate the performance of our algorithm.

The experiments are carried out using the Ising models and spin glasses on various of topologies, including $2$D lattices, random graphs, small world graphs, and complete graphs. Our results on error of free energies are compared against mean-field methods
including the na\"ive mean-field (NMF), Thouless-Anderson-Palmer equations (TAP), belief propagation (BP), and the neural-network-based variational autoregressive networks (VAN).
On the $2$D lattice without the external field, the graph is planar, so there are exact solutions~\cite{kac1952combinatorial}. Whereas on the other graphs, we adopt the exact (carefully designed) exponential algorithms~\cite{pan2019solving} (in a reasonable time) to compute exact free energy values for the evaluations.

The results are shown in Fig.~\ref{fig:relative error}. We can see that, in all experiments, our method outperforms all mean-field methods and the neural-network-based methods, to a large margin. In regular random graphs, small world networks, and the Sherrington-Kirkpatrick model, our accuracy is only limited by the machine precisions ($10^{-16}$). In the experiments we choose $\widehat {D}=50$ and $\widehat{\chi}=500$, and the computational time on each instance is of a few seconds.  
%Empirically, our method is faster than the mean-field methods which require converging the message passing equations, and the neural-network-based methods running on GPU. More results about the dependence of the bond dimensions, as well as the list of computational time, can be found at the supplementary materials.
Empirically, our method is faster than the mean-field methods and the neural-network-based methods. More results about the dependence of the bond dimensions and the computational time can be found in the Supplemental Material.
Moreover, it is worth noting that combining with the autodifferential for tensor networks~\cite{liao2019differentiable} immediately gives our method an ability to perform learning tasks using graphical models. 
In the Supplemental Material, we give an example of using our method to learn a generative model~\cite{lecun2015deep, kingma2013auto, dinh2014nice,dinh2016density,pmlr-v37-rezende15,JMLR:v17:16-272,pmlr-v48-oord16,goodfellow2014generative,ackley1985learning,lecun2012efficient,kingma2014adam} on hand-written digits of the MNIST dataset~\cite{MNIST}.

\paragraph{\label{sec:qc}Application to quantum circuit simulations--}
The problem of computing free energy of graphical models is similar to the problem of computing single amplitude estimates of a superconducting quantum circuit~\cite{boixo2018characterizing}, which can be treated as a graphical model with complex couplings.
Classical simulation of quantum circuits is important for verifying and evaluating the computational advances of quantum computers~\cite{markov2008simulating,chen2018classical,boixo2017simulation,PhysRevLett.123.190501,napp2019efficient,schutski2019adaptive}. However, the near-term noisy intermediate-scale quantum circuits (including Google's recently announced "supremacy circuit"~\cite{arute2019quantum}) are not perfect: each operation of them contains a small error. Thus, an important open question is whether approximate simulations of quantum circuits could beat the noisy quantum device. Answering this question apparently requires advanced studies of approximate algorithms for simulating quantum circuits.

\begin{figure}[h]
	\centering
	\includegraphics[width=0.48\columnwidth]{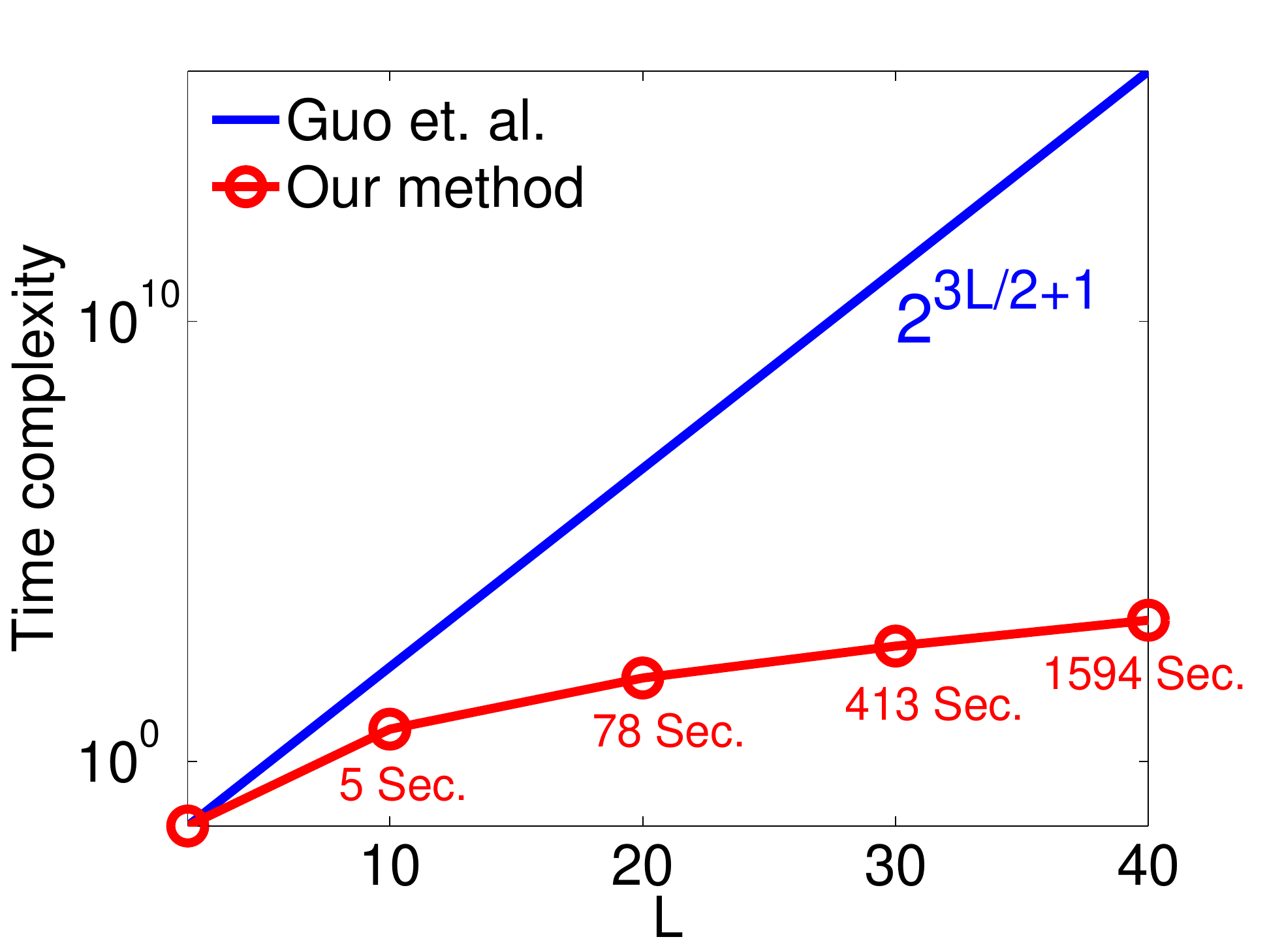}
	\includegraphics[width=0.48\columnwidth]{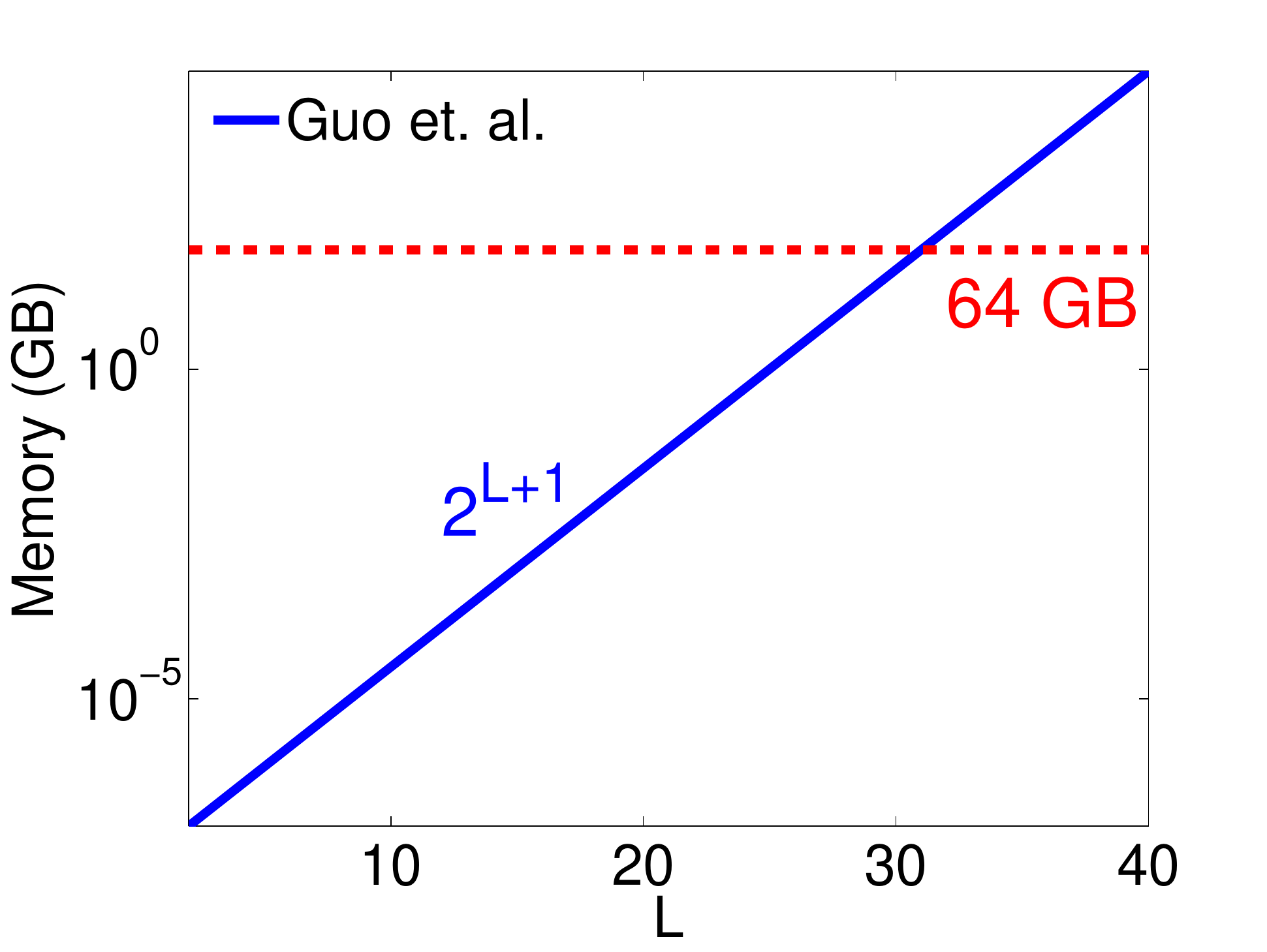}
	\caption{\label{fig:qc} Computational time and memory usage of our algorithm in simulating random quantum circuits with depth $d=8$, comparing with the exact tensor network method of Guo et al.~\cite{PhysRevLett.123.190501}. We ran our algorithm on a workstation with $64$ GB memory (as indicated by the red dashed line). The blue lines with formulas in the figure represent the precise time and space complexity of the exact algorithm~\cite{PhysRevLett.123.190501}. The memory usage is calculated based on double precision complex number. Each red point in the left panel is averaged over $10$ random circuits, the error bars are much smaller than the symbol size.}
\end{figure}

Our method directly applies to approximate single-amplitude simulation of quantum circuits with any kind of connectivities, such as two-dimensional lattice~\cite{boixo2017simulation,PhysRevLett.123.190501}, and random regular graphs as considered in the quantum approximate optimization algorithm~\cite{farhi2014quantum}, after converting the initial state, the measurement qubit string, and the gates into tensors. The key difference between our method and existing methods for quantum circuit simulation is that, by detecting low-rank structures in the circuit, our method heavily reduces the computational complexity. Although this introduces SVD truncation errors, we will illustrate that at least in the shallow circuits the error is almost negligible. We perform experiments using standard random circuits on two-dimensional lattices~\cite{chen2018classical,boixo2017simulation,PhysRevLett.123.190501}, which iteratively apply single-qubit gates and two-qubit controlled Z gates to the initial $|0,0,...,0\rangle$ state, and finally measure the amplitude of a specific qubit string. The generation protocol is described in detail in the Supplemental Material. We evaluate the performance of our method against the recently developed state-of-the-art exact tensor contraction method~\cite{PhysRevLett.123.190501}, which has a precisely predictable space and time complexity. 
%So in comparisons we do not have to run their algorithm which may require a supercomputer. 
%We run our algorithm on a workstation with $64$ Gigabytes memory. 
With depth $d=8$, our algorithm can handle circuits with at most $40\times 40=1600$ qubits with SVD accumulated truncation error $\epsilon_{\textrm{SVD}}\leq 10^{-12}$ on a workstation with $64$ GB memory in an hour. As compared in Fig.~\ref{fig:qc}, the computational complexity of our method is much lower than the method of~\cite{PhysRevLett.123.190501} . The right panel of Fig.~\ref{fig:qc} indicates that the method of ~\cite{PhysRevLett.123.190501} already costs at least $64$ GB memory for storing the largest intermediate tensor with $L=31$ and further requires $32$ TB memory for handling $L=40$.
%We note that when applied to deep circuits such as , instead
We note that so far our algorithm cannot handle the circuit with a large depth such as Google's circuit~\cite{arute2019quantum} with a small SVD error, because the current implementation of our algorithm only works on a single workstation: this prevents us from using a large bond dimension.

%\balance
\paragraph{Discussions--}
We have presented an algorithm for contracting arbitrary tensor networks, based on the matrix product state for automatic detecting of low-rank structures inside the tensor networks during the contraction process. We have demonstrated advances of our method in the inference and learning in graphical models and in simulation of shallow quantum circuits.
The particular strength of our method is able to find the internal low-entanglement structures automatically in the irregular tensor networks. 
%and show applications of our algorithm in the inference and learning in graphical models, and classical simulation of quantum circuits. 
%Based on the matrix product states, its canonical form, and efficient approximations using the singular value decompositions. 
%Our algorithm detects low-rank structures inside the tensor networks during the contraction process, hence reduces both time and space complexities. 
%Although we can not guarantee that our algorithm works accurately for tensor networks encoding strong entanglements, it indeed significantly outperforms existing methods in many applications. We have demonstrated advances of our method in the inference and learning in graphical models, and in simulation of shallow quantum circuits.
%In graphical models, we have demonstrated that our method gives much smaller error of the free energy for spin glass sytems on lattices, random graphs, and on complete graphs. In simulation of quantum circuits, we have shown that our algorithm has much smaller computational complexity than the state-of-the-art simulation methods, while producing only negligible truncation errors. 
%In this work we adopted simple greedy contraction orders. This can be improved using more sophisticated contraction ordering, we put this into future work. Moreover, 
The MPS representation of tensors in our method naturally supports distributed storage. It is interesting to see how large a quantum circuit we can simulate if a supercomputer is accessible to our algorithm.
Another interesting development is exploring learning with quantum circuits using our scheme and backpropagation.
We hope more advanced arbitrary tensor network contraction methods inspired by our approach could fully release the numerical computational power of tensor networks to wider applications in science and engineering.
A \textsc{PYTHON} implementation of our method is available at~\cite{code}.

%A \textit{python} implementation of our method and algorithm, together with the datasets used in our experiments, are available at~\cite{code}.

%\paragraph{Acknowledgements--}
\begin{acknowledgments}
	We thank Song Cheng, Haijun Liao, Chuang Wang, Lei Wang, Tao Xiang, Zhiyuan Xie, Haijun Zhou, and the BFS Tensor community for inspiring discussions, and Jinguo Liu for generating quantum circuits using \textit{Yao.jl}~\cite{luo2019yao}.
	%~\cite{luo2019yao}.
	%P.Z. is supported by project QYZDB-SSW-SYS032 of CAS, and Project 11947302 and 11975294 of NFSC.
	P.Z. is supported by project QYZDB-SSW-SYS032 of Chinese Academy of Sciences, and the National Natural Science Foundation of China under Grants No. 11947302 and No. 11975294.
\end{acknowledgments}

\bibliography{tn.bib}% Produces the bibliography via BibTex

\clearpage

\onecolumngrid
\appendix  
\balance

\section{\label{detailed contraction} 
	Detailed description of the contraction process}
\begin{figure*}[h]
	\includegraphics[width=\columnwidth]{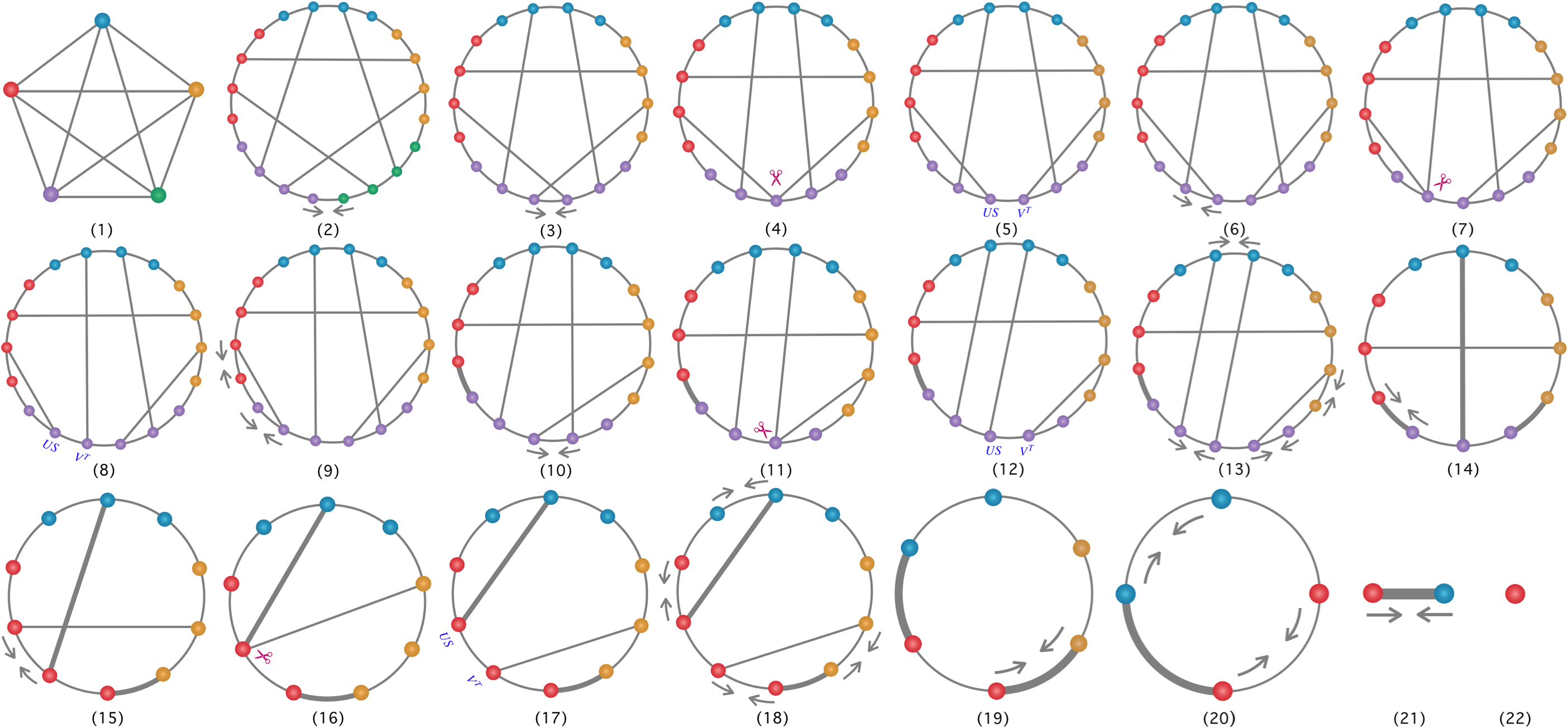}
	\caption{\label{fig:detail} Detailed process of contracting a tensor network with $5$ nodes, each of which is a four-way tensor, as sketched in Fig. 2 of main text. The scissor symbols in the figures indicate applying SVDs on the matrices unfolded from the tensors. The tensors in the step (1) are the original four-way tensors connected to each other, forming a fully connected pentagon. The step $(2)$ shows the MPS representation transformed from $(1)$; the arrow indicates \textit{contracting} two MPSes, that is, annihilating one color. As an example, the green MPS and the purple MPS is contracted to a longer purple MPS in $(3)$. The steps $(3)-(5)$ show the \textit{swap} operation between two tensors in the purple MPS. To accomplish swapping, we contract two purple tensors first, then apply the SVD on the contracted tensor as shown in $(3)$ and $(4)$. Note that in $(5)$, we keep the canonical form of the MPS. In steps $(6)-(8)$, the \textit{swap} operation is repeated until two tensors connecting the same pair of colors are switched to adjacent positions. In steps $(9)$ and $(10)$, we finish the \textit{merge} step by contracting two tensors indicated by the arrow, producing a thick bond between red and purple MPSes. The steps $(10)-(14)$ represent the \textit{merge} between the purple-and-brown MPS pair and the purple-and-blue MPS pair. The steps $(14)-(22)$ depict the procedure mentioned above repeatedly until a scalar left in the end of the whole contraction process. }
\end{figure*}

The pseudo code of the algorithm is listed in the Algorithm~\ref{algo}. In the Algorithm list, the connectivity of the tensor network is denoted by a graph $\mathcal{G}$, its vertex set is denoted by $\mathcal{V}$, and its edge set is denoted by $\mathcal{E}$; the notation $D_{ij}$ represents the dimension of the bond $(ij)$. We also give a simple example by contracting a tensor network composed of $5$ nodes, each of which is a four-way tensor connecting to each other, with the step-by-step contraction process illustrated in Fig.~\ref{fig:detail}. 
%The (2)-(14) steps expound the process of annihilating the green nodes, then (14)-(19) represent annihilating of the purple nodes. The steps (19)-(21) explain the contraction process of yellow and blue nodes and a red node survives finally in step (22). Notice that we accomplish \textit{swap} by using the SVD, and keep the canonical form of MPS. 
%Starting from the original tensor network with $5$ colors representing $5$ distinct tensors, we first convert each tensor to a matrix product state with the same color. Then at each step of contraction we pickup an edge connecting different colors and eliminate it by contracting corresponding two tensors. This annihilates a color. The process is repeated until there is only one color left, denoting the final contraction results.
\begin{algorithm}[H]
	\caption{\textit{MPS calculus}\label{algo}  }
	
	\begin{algorithmic}  
		\Require  
		Tensor networks with tensors $\mathcal{A}^{(1)}$... $\mathcal{A}^{(n)}$, and the connectivity graph $\mathcal {G}(\mathcal V,\mathcal E)$; the maximum physical bond dimension $\widehat D$, the maximum virtual bond dimension $\widehat\chi$.  
		\Ensure  
		Contraction result $Z$.\\
		%\For{l=1 : $|V|$}
		
		\State Convert every tensor to the MPS representation.
		\While{$|\mathcal{V}|>1$}
		%     \For{$(\mu,\nu)\in \mathcal{E}$}
		%       \State $\widetilde %D_{(\mu,\nu)}\leftarrow \sum_{b\in \partial \mu}\log(D_{b,\mu})+\sum_{b\in \partial \nu}\log(D_{b,\nu})-2\log(D_{\mu,\nu})$
		%     \EndFor
		%\State $(i,j)\leftarrow \argmin_{(\mu,\nu)\in\mathcal E}\left[ \sum_{b\in \partial\mu}\log(D_{b,\mu})+\sum_{b\in \partial\nu}\log(D_{b,\nu})-2\log(D_{\mu,\nu})\right]$
		\State Select an edge $(i,j)$ according to a contraction order.
		%\State $(i,j)\leftarrow %\argmin_{(\mu,\nu)}\widetilde %D(\mu,\nu)$,
		%\State Keep $\mathcal{A}^{(i)}$ in the left canonical form and  $\mathcal{A}^{(j)}$ in the right canonical form.
		\State Move the local tensor corresponding to the edge $(i,j)$ in the $\mathcal {A}^{(i)}$ to the tail position of the MPS representation.
		\State Move the local tensor corresponding to the edge $(i,j)$ in the $\mathcal {A}^{(j)}$ to the head position of the MPS representation.
		
		%\State $\mathcal{A}^{(i)}\leftarrow \mathcal{A}^{(i)}\times\mathcal{A}^{(j)}$.
		\State Merge two MPSes $\mathcal{A}^{(i)}$ and $\mathcal{A}^{(j)}$ by contracting the edge $(i,j)$ that connects them, resulting to new MPS $\mathcal{A}^{(i)}$; drop $\mathcal{A}^{(j)}$. 
		
		\State $\mathcal{E}\leftarrow \mathcal{E}\setminus\{(i,j)\}$
		
		\For{$k\in \partial j$}
		\State $\mathcal{E} \leftarrow \mathcal{E}\setminus\{ (j,k)\}$
		
		\If{$k\in \partial i$}
		\State Do \textit{swap} operations to move the duplicated $(i,k)$ edges to the adjacent positions in both MPSes $\mathcal{A}^{(i)}$ and $\mathcal{A}^{(k)}$.
		\State \textit{merge} the adjacent local tensors, so that the two edges are combined, with a larger bond dimension $D_{k,i}$.
		\If{$D_{k,i}>\widehat D$}
		\State Canonicalize $\mathcal A^{(i)}$ and $\mathcal A^{(k)}$. 
		\State Contract two tensors connected by the edge $(i,k)$. 
		\State Do SVD on the unfolded matrix of the obtained tensor, and perform truncation on singular values to reduce $D_{k,i}$ to $\widehat D$.
		\EndIf
		\Else
		\State  $\mathcal{E} \leftarrow \mathcal{E}\cup\{ (i,k)\}$
		\EndIf
		\EndFor
		\State $\mathcal{V}\leftarrow{\mathcal{V}}\setminus\{j\}$.
		\EndWhile
		
		\State  Return $Z=\mathcal{A}^{(i)}$
	\end{algorithmic} 
\end{algorithm}
%The operations
% \etextit{Cong scalartraction}, \textit{merge} and \textit{swap} during the contraction process are illustrated in great detail. 

\section{Contracting two MPSes to a single MPS}
As an example, consider merging two tensors $ {\mathbf{\mathcal{A}}}=\{a_{jik}\}$ and $\mathbf{ \mathcal {B}}=\{b_{\mu i \nu}\}$ 
into tensor $\mathbf{\mathcal {C}}=\{c_{jk\mu \nu}\}$, 
where tensors are all in the MPS representation formulated as
%\begin{align} a_{ijk}&=\sum_{\alpha}\sum_{\beta}a^{(1)}_{j\alpha}a^{(2)}_{\alpha i \beta}a^{(3)}_{\beta k},&\,\,\,\,\, b_{\mu i\nu}&=\sum_{\alpha}\sum_{\beta}b^{(1)}_{\mu\alpha}b^{(2)}_{\alpha i \beta}b^{(3)}_{\beta \nu}.  \end{align}
\begin{align} a_{ijk}&=\sum_{\alpha}\sum_{\beta}a^{(1)}_{j\alpha}a^{(2)}_{\alpha i \beta}a^{(3)}_{\beta k},\\
b_{\mu i\nu}&=\sum_{\alpha}\sum_{\beta}b^{(1)}_{\mu\alpha}b^{(2)}_{\alpha i \beta}b^{(3)}_{\beta \nu}.  \end{align}
The process is illustrated using tensor diagram notations in Fig.~\ref{fig:eat}.
To ensure that summing over the index $i$ results to another MPS, we first do \textit{swap} on $\mathbf{\mathcal{A}}$ to switch the indices $i$ and $k$
\begin{align} a_{ijk}&=\sum_{\alpha}a^{(1)}_{j\alpha}a^{(23)}_{\alpha i k}\approx \sum_{\alpha,\beta}a^{(1)}_{j\alpha}\widehat a^{(2)}_{\alpha k \beta}\widehat a^{(3)}_{\beta i}, \end{align}
%    $a_{ijk}=\sum_{\alpha}a^{(1)}_{j\alpha}a^{(23)}_{\alpha i k}\approx \sum_{\alpha,\beta}a^{(1)}_{j\alpha}\widehat a^{(2)}_{\alpha k \beta}\widehat a^{(3)}_{\beta i}$,
%\end{align}

\begin{figure}[tb]
	\centering
	\includegraphics[width=0.5\columnwidth]{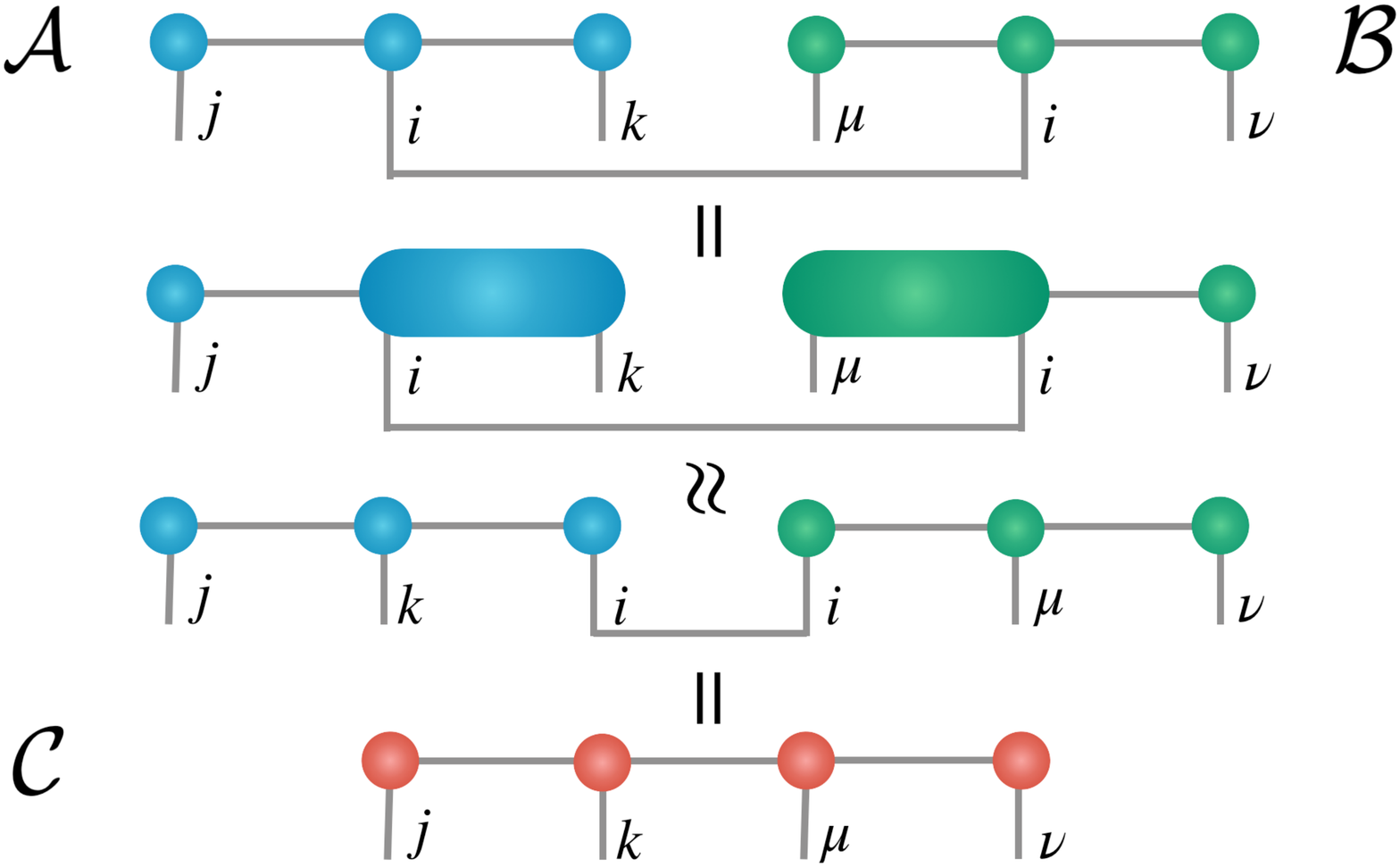}
	\caption{\label{fig:eat} Pictorial representation of the process of merging MPSes ${\mathbf{\mathcal{A}}}=\{a_{jik}\}$ and $\mathbf{ \mathcal {B}}=\{b_{\mu i \nu}\}$ 
		into another MPS $\mathbf{\mathcal {C}}=\{c_{jk\mu \nu}\}$.}
\end{figure}

where $a^{(23)}_{\alpha i k}$ are elements of the tensor created by contracting the index $\beta$; and the last step of the above equation requires the singular value decomposition, which could introduce truncations in the singular values. Similarly, we apply the \textit{swap} operation also on tensor $\mathbf{\mathcal{B}}$, to switch indices of $i$ and $\mu$, giving
\begin{align} b_{\mu i\nu}=\sum_{\beta}b^{(12)}_{\mu i\beta}b^{(3)}_{\beta \nu}\approx \sum_{\alpha,\beta}\widehat b^{(1)}_{i\alpha}\widehat b^{(2)}_{\alpha \mu \beta} b^{(3)}_{\beta \nu}.  \end{align}
%    $b_{\mu i\nu}=\sum_{\beta}b^{(12)}_{\mu i\beta}b^{(3)}_{\beta \nu}\approx \sum_{\alpha,\beta}\widehat b^{(1)}_{i\alpha}\widehat b^{(2)}_{\alpha \mu \beta} b^{(3)}_{\beta \nu}$.

After performing the swap operations on both tensors, we can see that the index $i$ locates at the tail position of the MPS representation of $\mathbf{\mathcal{A}}$ and at the head position of $\mathbf{\mathcal{B}}$. 
Thus summing over index $i$ results to a longer MPS $\mathbf{\mathcal{C}}$, as shown in the bottom Fig.~\ref{fig:eat}. 

\section{Canonical form of the MPS}
\begin{figure*}[htb]
	\centering
	\includegraphics[width=0.8\columnwidth]{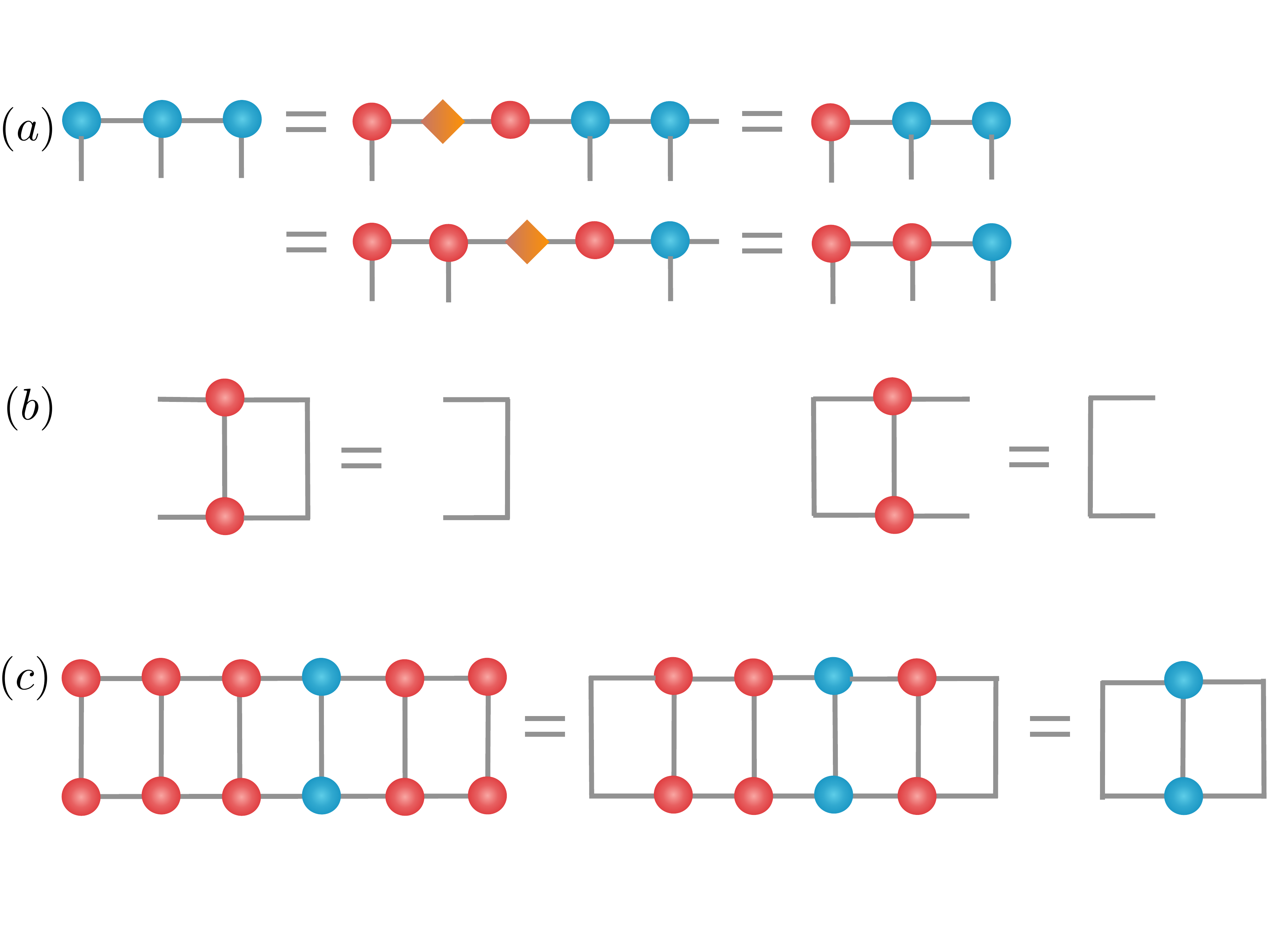}
	\caption{Illustration of the canonical form. (a) process of left canonicalization. (b) property of isometries. (c) norm of the whole MPS converts to the norm of single $3$-way tensor which is not an isometry. See descriptions in the text.}
	\label{fig:canonical}
\end{figure*}

When a tensor in dimension $2^n$ is represented as an MPS composed of three-way tensors $\{ \mathcal A^{(1)},\mathcal A^{(2)},\cdots,\mathcal A^{(n)} \}$, every element of the tensor can be written as product of matrices
$$a_{i_1,i_2,\cdots,i_n}=A_{i_1}^{(1)}\times A_{i_2}^{(2)}\times\cdots\times A_{i_n}^{(n)}.$$
First notice that inserting identity matrices does not change any thing, this suggests that there is a gauge degree of freedom which allows inserting any product of isometries $U$ and $U^\dagger$ into the above equation:
$$a_{i_1,i_2,\cdots,i_n}=A_{i_1}^{(1)}\times U\times U^\dagger \times A_{i_2}^{(2)}\times\cdots\times A_{i_n}^{(n)}.$$
This redundancy can be fully eliminated by the so-called canonical form~\cite{orus2014practical,schollwock2011density}, which forces all $3$-way tensors except one of them, say $i$-th tensor, to be isometry. If $i=n$ ($i=1$), that is in the right (left) most position, the form is known as \textit{left} (\textit{right}) canoincal position. Otherwise it is known as $\textit{mixed}$ canonical form. 

An example of converting an MPS to the \textit{left canonical} is illustrated in the top panel of Fig.~\ref{fig:canonical}. The blue tensors are not in isometry, while the red tensors are isometies, and the diamond represent diagonal matrix. The canonicalization is proceeded by \\
\indent1. Performing singular value decompositions on the left-most tensor, produce $A^{(1)} = USV^\top$ where $U$ and $V$ are isometries, and $S$ is a diagonal matrix storing singular values in the diagonal elements. \\
\indent2. Absorbing $S$ and $V$ by the $3$-way blue tensor in the middle, then perform singular value decompositions to the matrix unfolded by the tensor, producing isometries and diagonal tensors.\\
\indent3. Keep the absorbing and SVD to every tensor in order, until all tensors, except the last one become isometry.\\

In addition to eliminating the gauge degrees of freedom, the second advantage of the MPS is that the (global) norm of the whole MPS becomes the norm of the local $3$-way tensor which is the only one that is not an isometry. To show thismore clearly, consider performing partial trace over a $3$-way isometry tensor $\mathcal{A}$, as illustrated in Fig.~\ref{fig:canonical}, it follows that $$\sum_{\beta,\gamma}\mathcal{A}_{\alpha\beta\gamma}(\mathcal{A}^{\dagger})_{\gamma\beta\alpha'}=\delta^{\alpha}_{\alpha'}.$$ 
With this good property, in a canonical form, when all tensors except a certain one are isometry, computing the norm of the MPS, as well as computing sinple-point and two-point measurements, can be translated to a computation of single-tensor quantity. An example is shown at the bottom panel of Fig.~\ref{fig:canonical}, where we can see that after the reduction the norm of the whole MPS becomes the norm of a single $3$-way tensor.
To our algorithm, this property is important for us to do precise approximations during the \textit{merge} operation, where we need to do low-rank approximation between two connected MPSes by truncating the corresponding bond dimension, the canonical form transfers the local SVD truncations to a more global truncations involving the two MPSes, rather than involving only the two local tensors. 
We refer to~\cite{orus2014practical,schollwock2011density} for introductions of the matrix product states and the canonical form. 

%In this work, we apply the orthogonality $VV^{T}=\mathrm{I} $ of SVD operation to keep right canonical form of MPS for the left running, such as Fig.~\ref{fig:detail} (5), (8), (12) and (17). Otherwise, the left canonical form is similar.
%\section{\label{sec:equvalence} Proof of equivalence of BP and our method on tree graphs} 
%{\color{red} (obviously, our method is equivalent to variable elimination,  also famous of sum-product variable elimination. while  on a tree our method with MPS calculus doesn't need \textig{swap} operation  if with a  predefined  contraction order. And we know sum-product variable elimination differ from belief propagation that intermediate generated factors from initial factors  are  dropped one by one, but BP remain  all initial factors to compute all beliefs(essentially the intermediate generated tensors), finally obtaining all beliefs and marginals unchanged.  )}

\section{Contraction order}
Much effort is devoted to obtain efficient contraction algorithms on these tensor networks mostly based on renormalization group (RG) method to keep  topology as similar as the original regular lattices~\cite{white1992density,levin2007tensor,xie2009second,PhysRevB.86.045139}. However, when applied to problems out of physics such as graphical models,  the underlying network connections being random, far from lattices, methods based RG seem  to be in trouble. In essence, the key point is to find an optimal  contraction order to contract all edges to get a scalar in which way  computational memory and time is cost as little as possible. 
An important problem of the tensor network contraction is how to choose the edge order to eliminate one by one, which we refer to as \textit{contraction order}. 
In ~\cite{markov2008simulating}, Markov and Shi showed that optimal contraction sequences correspond to optimal (minimum width) tree decompositions of a tensor network’s line graph, relating the contraction sequence problem to a rich literature in structural graph theory. However finding the optimal tree decomposition for a general graph is a NP-hard problem, so usually one needs heuristic algorithms to find a good tree decomposition. Also notice that even equipped with the order given by the optimal tree decomposition, the algorithms for exact contraction in general is still an \textit{exponential algorithm} with computational complexity grows exponentially with the tree width of the line graph. In~\cite{dumitrescu2018benchmarking} tree decompositions of line graph of the tensor network are performed for finding minimum tree width; recently in~\cite{gray2020hyper} authors adopted the community-detection based methods for cutting the whole task to small tasks correponding to small communities, and in~\cite{huang2020classical} the authors adopted the stem optimizationm which tries to find the major component that takes a majority of compuatational cost and optimize accordingly.
Since we consider the approximate contraction using \textit{polynomial algorithms}, in contrast with the exact contraction, the dimension of the intermediate tensors in our scheme are hard to predict, because whether there are low-rank structures that we can use to reduce the dimensionalities is not known a priori. 

In this work, generally we adopt a greedy algorithm for sequentially selecting an edge from all remaining edges, which minimizes the dimension of the obtained tensor. 

That is
$$(i,j)\leftarrow \argmin_{(\mu,\nu)\in\mathcal E}\left[ \sum_{b\in \partial\mu}\log(D_{b,\mu})+\sum_{b\in \partial\nu}\log(D_{b,\nu})-2\log(D_{\mu,\nu})\right].$$
For some specific problems we have other choices. For example for 2-D lattice we could simply take a Zig-Zag order which respect the 2D regular structure of the tensor network.
Moreover, this can be improved using more sophisticated contraction ordering, for example using heuristic contraction orders given by tree decomposition of the line graph of the tensor network~\cite{markov2008simulating,chen2018classical,boixo2017simulation,PhysRevLett.123.190501}, or given by partition-based methods~\cite{gray2020hyper,huang2020classical}. 

\section{Dependences of the bond dimension \label{sec:experimental details}$\widehat{D}$ in the  graphical model experiments}
\begin{figure*}[htb]
	\includegraphics[width=\columnwidth]{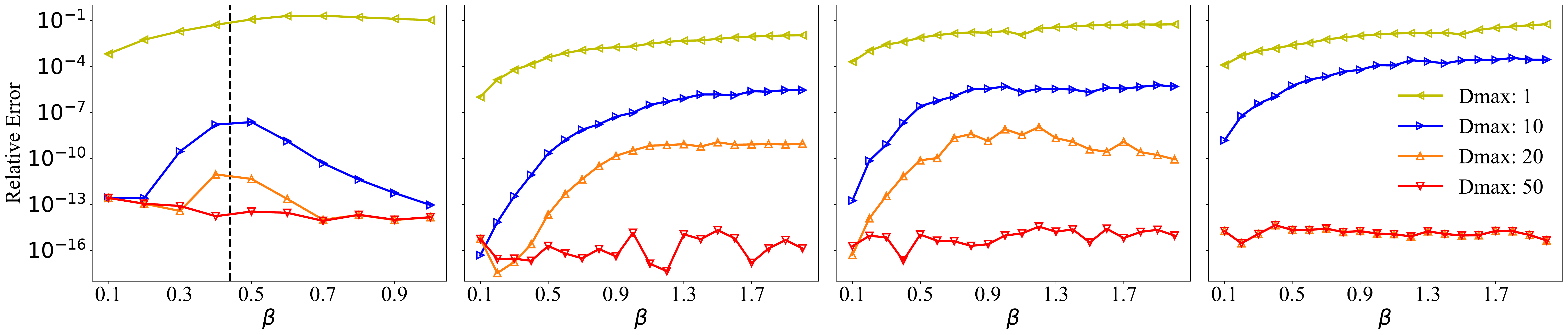}
	\caption{\label{fig:relative error tn} Relative error of free energy on different models with varies $\widehat D$ values, the experiment setting are identical to Fig. 3 of the main text.}
\end{figure*}

\begin{figure*}[htb]
	\includegraphics[width=\columnwidth]{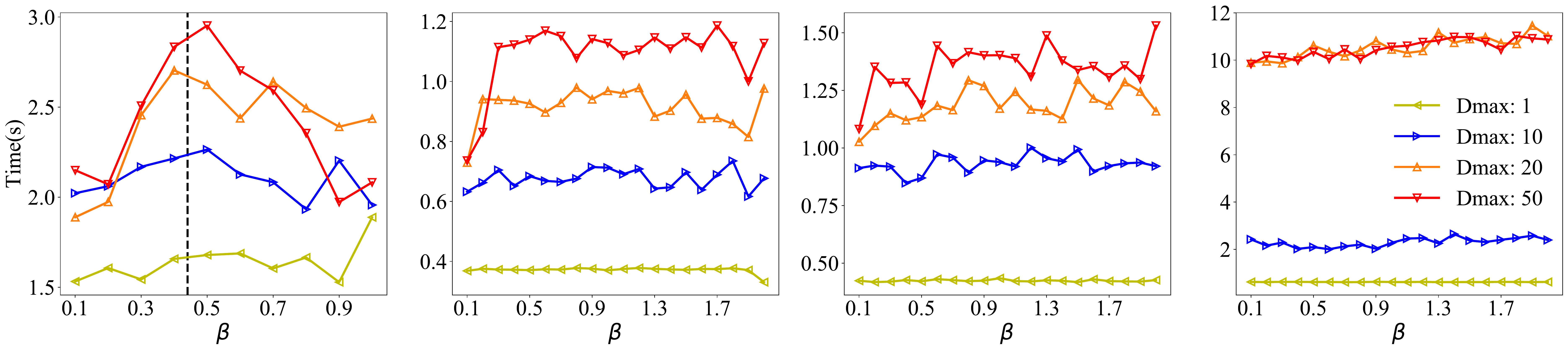}
	\caption{\label{fig:time tn} Time consumptions of the tensor network contraction algorithm in the free energy experiment of main text.}
\end{figure*}

In our experiments on graphical models, $\widehat{D}$ affects not only the running time but also the overall performance of the algorithm. In Fig.~\ref{fig:relative error tn} we show how the results are influenced by changing the maximum physical bond dimension $\widehat{D}$. The experimental settings are identical to the main text. 
First, as expected, it is clearly shown that the relative error becomes smaller as $\widehat{D}$ increasing for most situations. The only exception is with $\widehat{D} = 50$ on the $20$-spin SK model, where the results have no difference compared to $\widehat{D} = 20$. This means $\widehat{D} = 20$ is large enough to deal with the $20$-spin SK model at the $\beta$ range considered here.
In Fig.~\ref{fig:time tn}, time usage of our algorithm with different $\widehat D$ are shown. Since a bigger $\widehat{D}$ results to larger tensors, the time usage naturally increases. But for some cases, a bigger $\widehat{D}$ occasionally changes the contraction order and avoid some approximation operations, leading to a lower running time than a smaller $\widehat{D}$.

\section{Comparison between our method and the High Order Tensor Renormalization Group}
\begin{figure}[htb]
	\centering
	\includegraphics[width=0.8\linewidth]{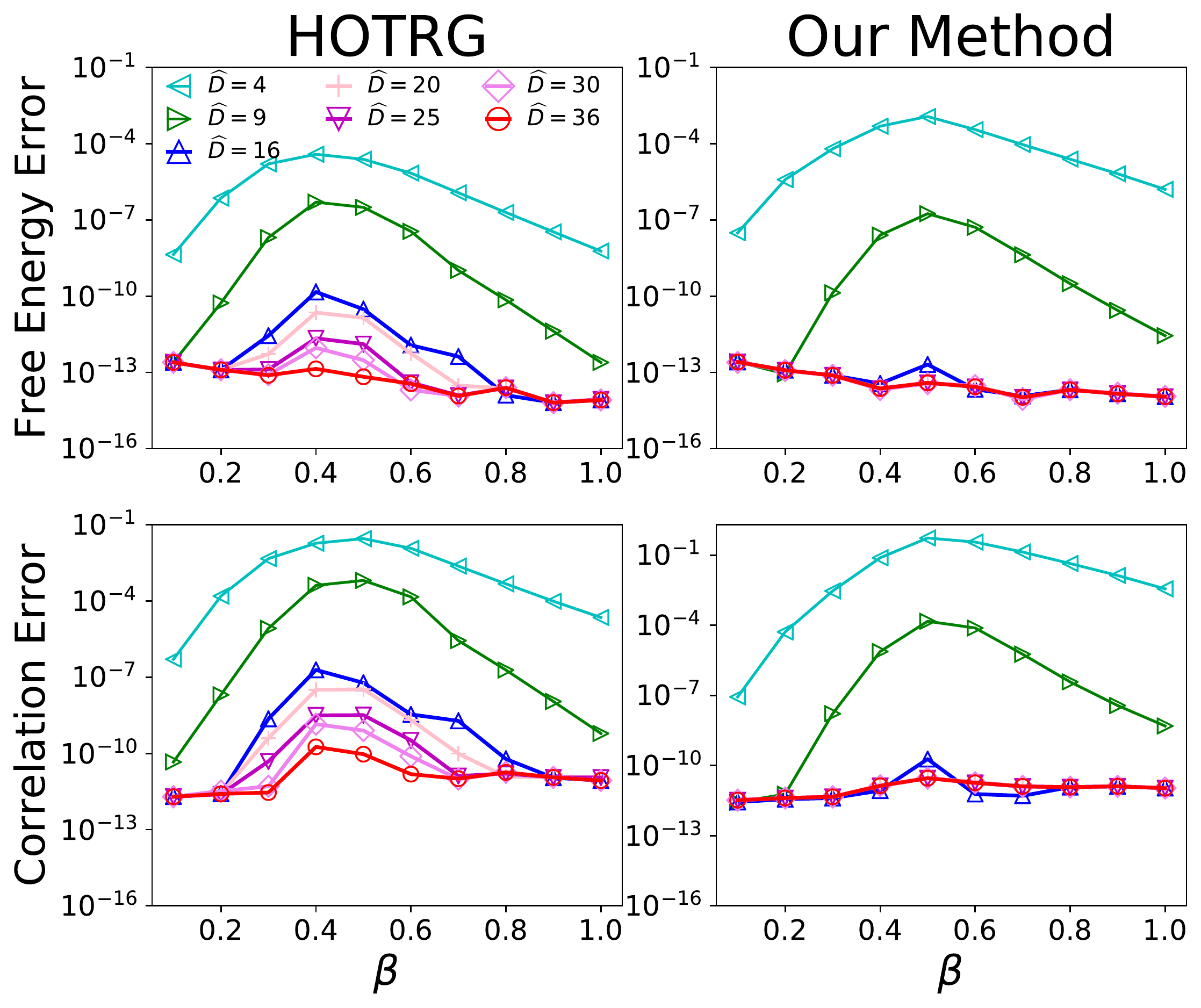}
	\caption{Relative error of free energy (\textit{Top}) and correlations (\textit{Bottom}) to exact solutions, of the High Order Tensor Renormalization Group (HOTRG)~\cite{PhysRevB.86.045139} and our method, on the ferromagnetic Ising model on $16\times 16$ square lattice with different inverse temperature $\beta$.
		Both HOTRG and our method have the same maximum physical bound dimension $\widehat D= 4, 9, 16, 20, 25, 30, 36$ from top to bottom.
		\label{fig:compare}
	}
\end{figure}

In this section we compare with the High Order Tensor Renormalization Group (HOTRG)~\cite{PhysRevB.86.045139} method in terms of reproducing both free energy and correlation functions. 
The HOTRG is one of the representative tensor renormalization group method, and has been shown to outperform other TRG methods such as the Levin-Nave Tensor Renormalization Group~\cite{levin2007tensor} and the Anisotropic Tensor Renormalization Group~\cite{adachi2019anisotropic}. 
The results are shown in Fig.~\ref{fig:compare}. In the figure we have compared the relative error of free energy $\Delta F = -\frac{1}{\beta L^2}\left|\ln(Z)-\ln(Z^{\text{exact}})\right|,$ and relative error of correlation functions $\Delta C = \sqrt{\sum_{(ij)}\left (C_{(ij)}-C_{(ij)}^{\text{exact}}\right)^2},$ on the ferromagnetic Ising model on a $L\times L$ lattice with $L=16$. The HOTRG stores many $4$-way tensors with dimension $\widehat D^4$ during the contraction process, and our method stores many $3$-way tensors with dimension $\widehat D\chi^2$. For a fair comparison, we force two methods to have the same space complexity, hence in our method we limit $\chi=\widehat D^{\frac{3}{2}}$. Moreover since comparisons are carried out only on $2$-D lattices (because the HOTRG can not be applied to irregular lattices), rather than the general greedy contraction order, we choose to use a simple Zig-Zag contraction order, which selects tensors row by row.

From the figure we can see that with the same bond dinmension $\widehat D$, our method works better than the HOTRG by giving smaller both free energy error and correlation error. In particular, with $\widehat D\geq 16$, in our method all error curves converge to almost machine precision, while the HOTRG requires $\widehat D\geq 36$ to reach a similar error.

\section{\label{sec:learning} Learning of graphical model using tensor networks}
Generative learning in the unsupervised learning models the joint distribution of random variables in the given data and generates new samples from the learnt distribution. It is an important task in modern maching learning~\cite{lecun2015deep} and find wide applications in many areas of artificial intelligence. Fashion generative models include variational autoencoders (VAE)~\cite{kingma2013auto}, normalizing flows~\cite{dinh2014nice,dinh2016density,pmlr-v37-rezende15}, autoregressive models~\cite{JMLR:v17:16-272,pmlr-v48-oord16} and generative adversarial networks (GANs)~\cite{goodfellow2014generative}. Here, we focus on a classical generative model known as the Boltzmann machine~\cite{ackley1985learning} with no hidden variables, which is also known as the inverse Ising model which is the maximum entropy model given the pairwise measurement of data. Using this example we demonstrate that our method for contracting the Ising model can be directly used for learning tasks. The objective function of the learning using an Ising model is the the negative log-likelihood $\mathcal L$, which we aim to minimize:
\begin{align}
\label{eq:loss}
\mathcal{L} & = -\frac{\log P(\mathbf{X})}{N} = -\frac{\log \prod_i P(\mathbf{x}_i)}{N}\\ \notag
&=-\frac{\beta}{N} \sum\limits_i\biggl( \sum\limits_{(mn) \in \mathcal{E}} J_{mn}x_{i,m}x_{i,n} + \sum\limits_m h_m x_{i,m} \biggr)+\log Z
\end{align}
In the last equation, $\mathbf{X}$ is the dataset, $\mathbf{x}_i$ is the $i$-th data, $N$ is the size of dataset and $\mathcal{E}$ represents edges of graphical model we employ.
In classical machine learning method, the partition function $\log Z$ appearing in the log-likelihood is difficult to compute, and people usually use approximated method such as the \textit{contrastive divergence}. Fortunately our method provides a relatively fast and accurate way to calculate $\log z$.
%, as well as its derivatives with respect to the model parameters via the \textit{back propagation} algorithm. 
Essentially, by setting the derivative of $\mathcal{L}$ with parameters $J_{mn}$ (couplings) and $h_m$ (external fields) to be zero, we can get  
%\begin{equation}
%\frac{\partial{\mathcal{L}}}{\partial{J_{mn}}} = %- \frac{\beta}{N}\sum\limits_ix_{i,m}x_{i,n} + %\frac{\partial{\log Z}}{\partial{J_{mn}}}
%\end{equation}

%\begin{equation}
%\frac{\partial{\mathcal{L}}}{\partial{h_m}} = - %\frac{\beta}{N}\sum\limits_ix_{i,m} + %\frac{\partial{\log Z}}{\partial{h_m}}
%\end{equation}
\begin{align}
\label{eq:moment}
&\beta C_{data}= \frac{\beta}{N}\sum\limits_ix_{i,m}x_{i,n} =\beta C_{model}= \frac{\partial{\log Z}}{\partial{J_{mn}}}\notag\\
&\beta M_{data}=\frac{\beta}{N}\sum\limits_ix_{i,m} =\beta M_{model} =\frac{\partial{\log Z}}{\partial{h_m}}.
\end{align}
The model parameters $\{J_{mn}\}$ and $\{h_m\}$ can be learnt by matching the moments of the model with the moments of data. We emphasis that, here we do not even need to calculate the correlations and magnetizations, because the gradient on $J_{mn}$ and $h_m$ can be estimated directly by taking derivative of the loss function by using the back-propagation algorithm. Then the learning can be carried out by utilizing a modern deep learning optimizer such as the stochastic gradient descent (SGD)~\cite{lecun2012efficient} and the ADAM~\cite{kingma2014adam} to update the parameters.

As a demonstration, we perform experiments on the handwritten digits of the MNIST dataset~\cite{MNIST} to show how to learn an Ising model from data using our TN method combined with the back-propagation algorithm. For preparation, we reshape $28\times28$ binarized images to $14\times14$ for faster contractions. Our graphical model is based on $2$D square lattice with additional diagonal connections and second nearest neighbors connections. As an demonstration, we use only first five images of MNIST as training set for learning the model.

\begin{figure}[htb]
	\includegraphics[width=\columnwidth]{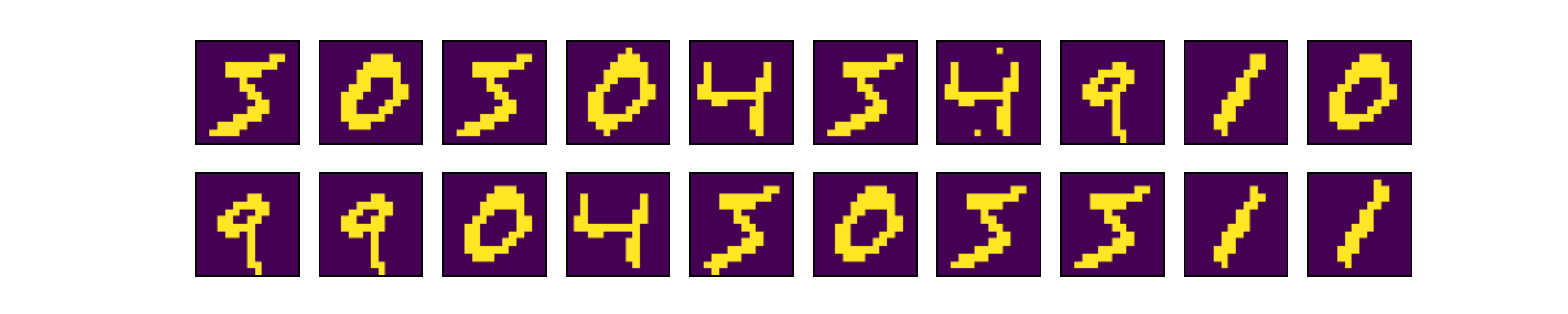}
	\caption{\label{fig:mnist} MCMC samples of Ising model learnt from $5$ handwritten images of the MNIST dataset.}
\end{figure}

After training through stochastic gradient descent, the Ising model displays the similar distribution as the empirical distribution of the training data, and  we can generate images by sampling from the distribution that our model has learned. Here we adopt traditional Markov Chain Monte Carlo (MCMC) to sample from the model, the samples are shown in Fig.~\ref{fig:mnist}. We can see that the images are well presented and similar to the training images. The negative loglikehood obtained is $2.41$ which is very close to the lower bound $\ln 5 = 1.61$. 

\section{\label{sec:qc stucture} Protocol for generating random quantum circuits}
\begin{figure}[htb]
	\includegraphics[width=\columnwidth]{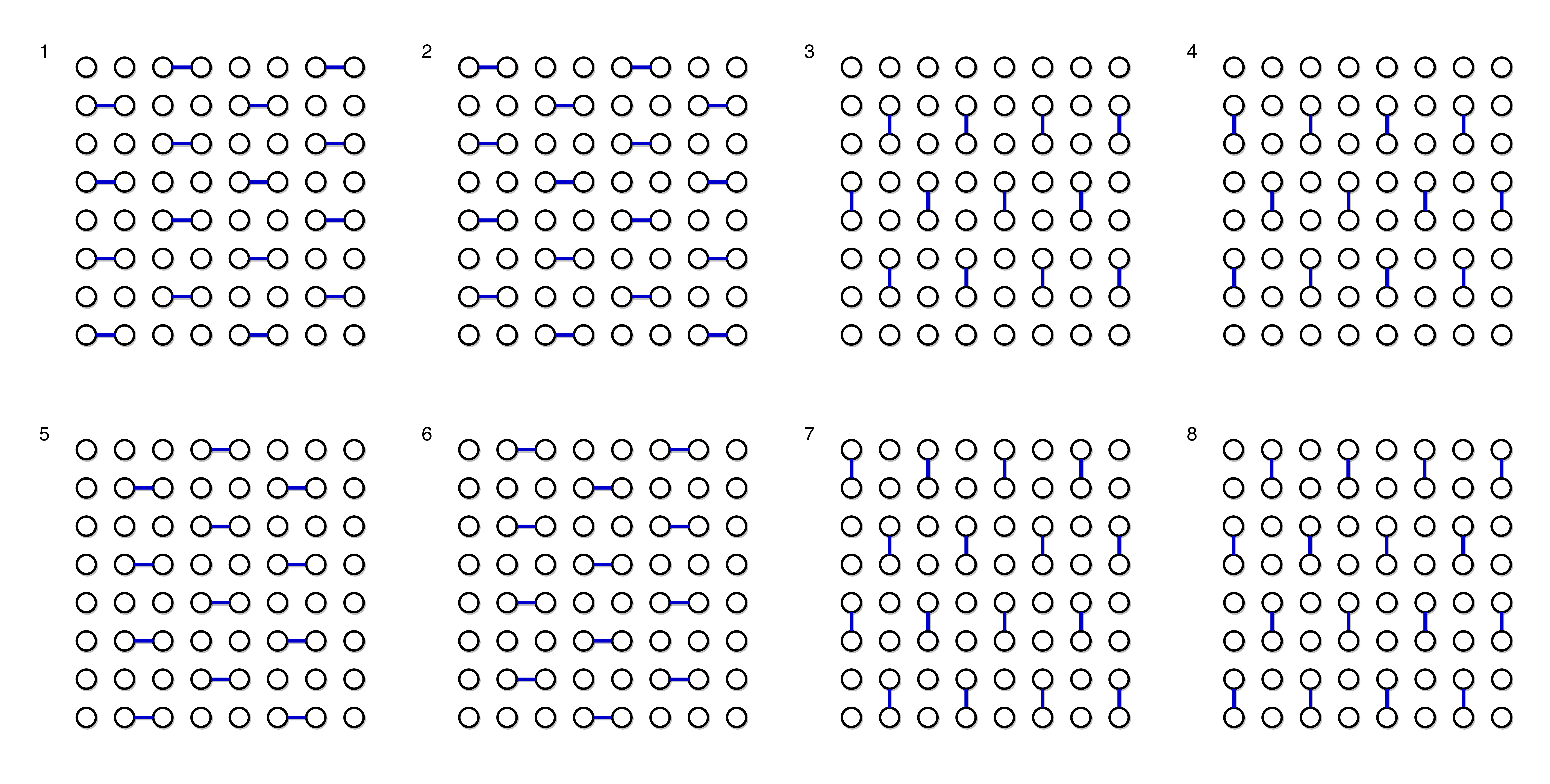}
	\caption{\label{fig:CZ} Choices of the two-qubit-gate layers in generating random quantum circuits.}
\end{figure}

The random quantum circuits with depth $d$ used in our experiments are generated as follows:
\begin{itemize}
	\item[1.] Apply a Hadamard gate to each qubit.
	\item[2.] Apply controlled-Z gates organized in one of the eight layouts as shown in Fig.~\ref{fig:CZ} 
	once a time alternatively, then apply a randomly chosen gate from $\{T,X^{1/2},Y^{1/2}\}$ to each qubit which is not acted by the CZ gates. 
	\item[3.] Repeat steps $2$ for $d-1$ times.
	\item[4.] Apply a Hadamard gate to each qubit. 
\end{itemize}
%
%\bibliography{tn.bib}% Produces the bibliography via BibTex

\end{document}